# Relativistic Electron Beam Acceleration by Compton Scattering of Lower-Hybrid Waves

R. Sugaya, T. Maehara and M. Sugawa

*Department of Physics, Faculty of Science, Ehime University, Matsuyama 790-8577, Japan*

**Abstract.** It has been proved theoretically and numerically that the highly relativistic electron beam can be accelerated efficiently via the Compton scattering induced by nonlinear Landau and cyclotron damping of the lower-hybrid waves.

## 1. INTRODUCTION

Acceleration and heating of a relativistic electron beam by the Compton scattering of electrostatic waves propagating almost perpendicularly in a magnetized plasma are investigated theoretically and numerically on the basis of the relativistic kinetic wave and transport equations[1-5]. Two electrostatic waves interact nonlinearly with the relativistic electron beam, satisfying the resonance condition for the Compton scattering (nonlinear Landau and cyclotron damping)[4,5],

$$\omega_k - \omega_{k'} - (k_\parallel - k'_\parallel)v_\parallel = m\omega_{ce} \quad , \tag{1}$$

where $v_\parallel \simeq v_b$, $v_b$ is the parallel velocity of the relativistic electron beam, $\omega_{ce} = eB_0/\gamma_b m_e c$ ($\gamma_b = (1 + p^2/m_e^2 c^2)^{1/2}$, $\boldsymbol{p} = \gamma_b m_e \boldsymbol{v}$) is the relativistic electron cyclotron frequency, and $m$ is an integer. The relativistic transport equations using the relativistic drifted Maxwellian momentum distribution function of the relativistic electron beam were derived and analyzed.

## 2. BASIC EQUATIONS

The relativistic kinetic wave equations of the two electrostatic waves and the relativistic transport equations for the relativistic electron beam are expressed as[5]

$$\frac{\partial U_k}{\partial t} = -A_0 U_k U_{k'} \quad , \quad \frac{\partial U_{k'}}{\partial t} = A_1 U_k U_{k'} \quad , \tag{2}$$

$$\frac{\partial U_b}{\partial t} = \frac{\omega_{k''}}{\omega_k} A_0 U_k U_{k'} \quad , \quad \frac{\partial \boldsymbol{P}_b}{\partial t} = \frac{\boldsymbol{k}''}{\omega_k} A_0 U_k U_{k'} \quad , \tag{3}$$

where $U_k = \Gamma_k |E_k|^2$, $\Gamma_k = \dfrac{1}{8\pi}\left[\dfrac{\partial(\varepsilon'_k \omega_k)}{\partial \omega_k}\right]$, $U_b = \int d\boldsymbol{p}\, n_b \gamma_b m_e c^2 g_b$, $\boldsymbol{P}_b = \int d\boldsymbol{p}\, n_b \boldsymbol{p} g_b$, $A_0 = -\dfrac{\omega_k}{4\pi \Gamma_k \Gamma_{k'}} A_{k,k'',k'}$, $A_1 = \dfrac{\omega_{k'}}{\omega_k} A_0$, $A_{k,k'',k'} = \text{Im}\left(C^{(b)}_{k,k'',k'} + D^{(b)}_{k,k'',k'}\right)$. Here, $U_k$ is the wave energy density, $\boldsymbol{E}_k = (\boldsymbol{k}/k)E_k$ is the wave electric field, the background stationary uniform magnetic field $\boldsymbol{B}_0 = (0, 0, B_0)$ is in the z-direction, $\omega_{k''} = \omega_k - \omega_{k'}$, $\boldsymbol{k}'' = \boldsymbol{k} - \boldsymbol{k}' = (k''_\perp, 0, k''_\parallel)$, $\boldsymbol{k} = (k_\perp, 0, k_\parallel)$ and $\boldsymbol{k}' = (k'_\perp, 0, k'_\parallel)$ are in the x-z plane, the linear damping rate of the electrostatic waves is assumed to be zero ($\gamma_k = \gamma_{k'} = 0$), $\varepsilon_k$ is the dielectric constant, $U_b$ and $\boldsymbol{P}_b$ are the energy and momentum densities of the relativistic electron beam, and $g_b$ is the momentum distribution function of the relativistic electron beam. The matrix elements $C^{(b)}_{k,k'',k'}$ and $D^{(b)}_{k,k'',k'}$ giving the nonlinear wave-particle coupling coefficients $A_0$ and $A_1$ are described in Ref. 5 in detail, where it is needed to set $v_d = \boldsymbol{E}_0 = 0$ in the matrix elements, because the cross-field drift velocity and the cross-field electric field are absent. Equations (2) and (3) yield the conservation laws for the total energy and momentum densities of the electrostatic waves and the relativistic electron beam.

Next we assume the momentum distribution function of the relativistic electron beam given by a relativistic drifted Maxwellian momentum distribution function,

$$g_b = \frac{\mu}{4\pi m_e^3 c^3 \beta K_2(\mu)} \exp\left[-\mu\beta\gamma_b\left(1 - \frac{v_b v_z}{c^2}\right)\right] \quad , \tag{4}$$

where $\beta = \left(1 - v_b^2/c^2\right)^{-1/2}$, $\mu = m_e c^2 / k_B T_b$ with the beam temperature $T_b$, $K_r$ is a modified Bessel function of the rth order. In particular, it is easily found that $g_b$ in the nonrelativistic limit ($\beta \simeq 1$, $\mu \gg 1$) becomes a usual drifted Maxwellian distribution function[2], $g_b = (2\pi m_e k_B T_b)^{-3/2} \exp\left[-m_e\left(v_x^2 + v_y^2 + (v_z - v_b)^2\right)/2k_B T_b\right]$. Thus the energy and momentum densities of the relativistic electron beam become as follows:

$$U_b = n_b m_e c^2 \beta \left(\frac{K_3(\mu)}{K_2(\mu)} - \frac{1}{\mu\beta^2}\right) \quad , \quad P_{bz} = n_b m_e v_b \beta \frac{K_3(\mu)}{K_2(\mu)} \quad , \quad P_{bx} = P_{by} = 0 \quad . \tag{5}$$

Here, it is found that $U_b$ and $P_{bz}$ in the nonrelativistic limit are reduced to the usual forms expressed by $U_b = n_b\left(m_e c^2 + m_e v_b^2/2 + 3k_B T_b/2\right)$ and $P_{bz} = n_b m_e v_b$.

We investigate the acceleration and heating of the relativistic electron beam with the nonrelativistic beam temperature of $\mu \gg 1$. By means of the asymptotic expansion of $K_r(\mu) = (\pi/2\mu)^{1/2} e^{-\mu}\left[1 + (4r^2 - 1)/8\mu\right]$, $U_b$ and $P_b$ can be approximated as $U_b = n_b m_e c^2 \beta\left(1 + \frac{5}{2\mu} - \frac{1}{\mu\beta^2}\right)$ and $P_{bz} = n_b m_e v_b \beta\left(1 + \frac{5}{2\mu}\right)$. Immediately we can get the simple relativistic transport equations with $\beta, \mu \gg 1$ from Eqs. (3), and they are expressed in the followings:

$$\frac{\partial}{\partial t}\left(n_b m_e c^2 \beta\right) = \frac{\omega_{k''}}{\omega_k} A_0 U_k U_{k'} \quad , \tag{6}$$

$$\frac{\partial}{\partial t}\left(n_b k_B T_b\right) = \left[\frac{k''_\parallel c^2 - \omega_{k''} v_b}{\omega_k v_b} - \frac{\omega_{k''}}{\beta^2 \omega_k}\right]\beta A_0 U_k U_{k'} \quad . \tag{7}$$

Equation (6) shows the acceleration of the relativistic electron beam, and Eq. (7) shows its heating and cooling. It can be proved from Ref. 5 that the following relation holds:

$$A_0, A_1 \propto \eta_m = \frac{k_\parallel''(v_{\parallel 0} - v_b) + m\omega_{ce}}{\left|1 - v_{\parallel 0}\omega_{k''}/k_\parallel'' c^2\right|\omega_{ce}} \quad , \tag{8}$$

where $v_{\parallel 0} = (\omega_{k''} - m\omega_{ce})/k_\parallel''$. In the case of $m = 0$, Eq. (8) becomes $A_0, A_1 \propto \eta_m = k_\parallel''(v_{\parallel 0} - v_b)\gamma_p^2/\omega_{ce}$ ($\gamma_p = (1 - v_{\parallel 0}^2/c^2)^{-1/2}$), and hence it can be stated that when $v_{\parallel 0} > v_b$ ($v_{\parallel 0} < v_b$) and $k_\parallel'' > 0$, the relativistic electron beam can be accelerated (decelerated) via $m = 0$ scattering[4,5]. For $m \neq 0$, we find that the relativistic electron beam can be accelerated (decelerated) always when $m > 0$ ($m < 0$), because of $\eta_m \simeq m\,|\,k_\parallel'' c/m\,|$ ($v_{\parallel 0} \simeq v_b \simeq c$)[4,5].

## 3. NUMERICAL ANALYSIS FOR LOWER-HYBRID WAVES

In order to investigate the detailed behavior of the Compton scattering of the two lower-hybrid waves, we performed the numerical analysis of the dimensionless nonlinear wave-particle coupling coefficients $\alpha_0 = (\varepsilon_b/\omega_{ci})A_0$ and $\alpha_1 = (\varepsilon_b/\omega_{ci})A_1$ for $m = 0, 1$, where $\varepsilon_b = n_b m_e c^2$, $\omega_{ci} = Z_i eB_0/m_i c$, $\omega_0 = \omega_k$, $\omega_1 = \omega_{k'}$, $\omega'' = \omega_{k''}$, $\mathbf{k}_0 = \mathbf{k}$, $\mathbf{k}_1 = \mathbf{k}'$. The numerical calculation was carried out under the plasma parameters of $\omega_{pi}^2/\omega_{ci}^2 = 500$, $\omega_{pb}^2/\omega_{ce0}^2 = 0.01$, $v_{te}/c = 0.05$, $m_i/m_e = 1840$, $Z_i = 1$, $T_e/T_i = 1$, $k_{\parallel 0} v_{ti}/\omega_{ci} = 0.03$, and $k_{\parallel 1} v_{ti}/\omega_{ci} = 0.028$. Here, $v_{ts} = (2k_B T_s/m_s)^{1/2}$ (s=e,i), and $\omega_{ce0} = eB_0/m_e c$ is the nonrelativistic electron cyclotron frequency. Furthermore, it was confirmed that $\left|\text{Im}\,D_{k,k'',k'}^{(b)}/C_{k,k'',k'}^{(b)}\right| \ll 1$, that is, the plasma shielding effect is negligibly small compared with the Compton scattering. Figure 1(a) exhibits $\alpha_0$ and $\alpha_1$ versus $\beta$ for $\gamma_p/\beta = 1.02$, $\mu = 1000$, $m = 0$, $\omega_0/\omega_{ci} = 26.46$, $k_{\perp 0}v_{ti}/\omega_{ci} = 1.6$, $\omega_1/\omega_{ci} = 23.75$ and $k_{\perp 1}v_{ti}/\omega_{ci} = 1.83$. The solid and dotted curves correspond to $\alpha_0$ and $\alpha_1$, respectively. Figure 1(b) exhibits $\alpha_0$ and $\alpha_1$ versus $\beta$ for $\gamma_p/\beta = 1$, $\mu = 1000$, $m = 1$, $\omega_0/\omega_{ci} = 24.39$, $k_{\perp 0}v_{ti}/\omega_{ci} = 1.8$, $\omega_1/\omega_{ci} = 20.83 \sim 22.67$ and $k_{\perp 1}v_{ti}/\omega_{ci} = 2.17 \sim 3.76$. The resonance condition Eq. (1) can be satisfied when $\omega_{ce} = \omega_{ce0}/\beta \lesssim \omega'' \approx 2\omega_{ci}$ with $\beta \gtrsim 1000$. In Fig. 2(a), the absolute value of $\alpha_0$ is shown versus $\gamma_p/\beta$ for $\beta = 2000$, $\mu = 100$, $m = 0$, $\omega_0/\omega_{ci} = 25.46$, $k_{\perp 0}v_{ti}/\omega_{ci} = 1.6$, $\omega_1/\omega_{ci} = 23.75$ and $k_{\perp 1}v_{ti}/\omega_{ci} = 1.83$. The solid and dotted curves correspond to $\alpha_0 < 0$ and $\alpha_0 < 0$, respectively. In Fig. 2(b), $\alpha_0$ is shown versus $\gamma_p/\beta$ for $\beta = 20000$, $\mu = 100$, $m = 1$, $\omega_0/\omega_{ci} = 25.46$, $k_{\perp 0}v_{ti}/\omega_{ci} = 1.6$, $\omega_1/\omega_{ci} = 23.65 \sim 23.72$ and $k_{\perp 1}v_{ti}/\omega_{ci} = 1.83 \sim 1.86$.

## 4. CONCLUSION

It can be verified that the Compton scattering of the lower-hybrid waves can accelerate efficiently the highly relativistic electron beam[6]. It can be available usefully to the acceleration of the highly relativistic electron beam. 

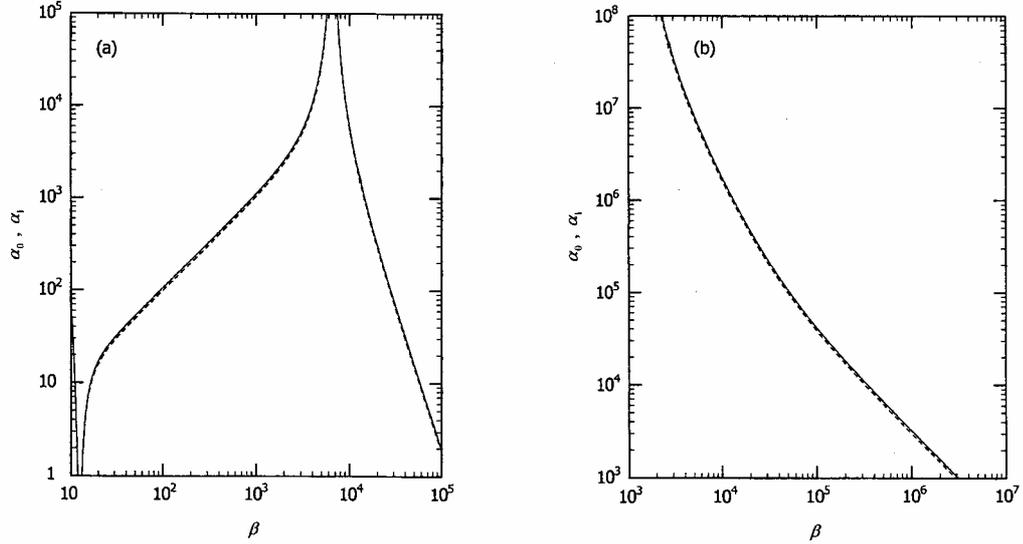

Fig.1. Here, $\alpha_0$ and $\alpha_1$ versus $\beta$ are shown.

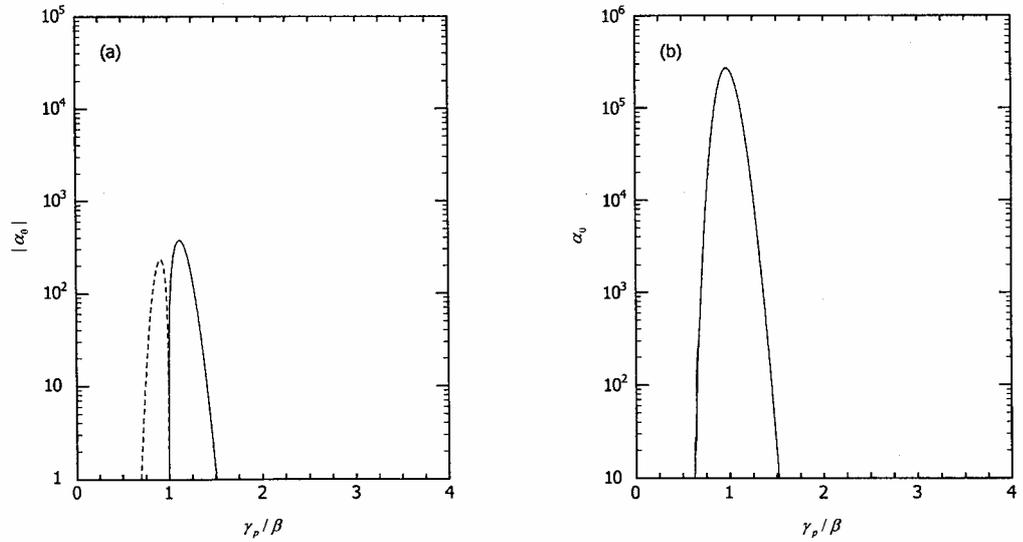

Fig. 2. Here, $|\alpha_0|$ and $\alpha_0$ versus $\gamma_p/\beta$ are shown.